\documentclass[11pt,twoside]{article}

\usepackage{asp2006}
\usepackage{fancyhdr,graphicx,caption,rotating,multirow,lscape}
\DeclareGraphicsExtensions{.eps,.eps.gz,.ps,.jpg,.tiff}

\begin{document}

\title{Searching for Planetary Transits in the Lupus Galactic Plane}  



\author{David T. F. Weldrake}
\affil{Max-Planck Instit\"ut f\"ur Astronomie, K\"onigstuhl 17, Heidelberg, 69117 Germany}
\author{Daniel D. R. Bayliss, Penny D. Sackett $\&$ Michael Bessell}  
\affil{Research School of Astronomy and Astrophysics, The Australian National University, Mount Stromlo, Cotter Road, Weston Creek, ACT 2611 Australia}    
\author{Brandon Tingley}
\affil{Institut d'Astronomy et Astrophysique, Universite Libre de Bruxelles,
B-1050 Brussels, Belgium}

\begin{abstract}
A $52' \times 52'$ field in the Lupus Galactic plane was observed with the ANU 1m telescope for 53 nights during 2005 and 2006 in a search for transiting Hot Jupiter planets. A total of 2200 images were obtained. We have sampled 120,000 stars via differential photometry, of which $\sim$26,000 have sufficient photometric accuracy ($\le 2.5 \%$) with which to perform a search for transiting planets. Ongoing analysis has led to the identification of three candidates. We present an overview of the project, including the results of radial velocity analysis performed on the first candidate (Lupus-TR-1) with the 4m AAT telescope. The third candidate, Lupus-TR-3 (P=3.914d, V$\sim$16.5), is a particularly strong case for a giant planet of 1.0-1.2R$_{\rm{Jup}}$ orbiting a solar-like primary star with a near central transit. Further observations are planned to determine its nature.
\end{abstract}

\section{Project Overview}
A deep, wide-field survey for transiting giant planets has been performed in the Lupus Galactic plane. Using the ANU 1m telescope, a 52$'\times$52$'$ field was observed for 53 nights in 2005 and 2006. A total of 2200 images were obtained with a wide passband V+R filter. Via an application of differential photometry, light curves have been constructed for a total of 120,000 stars (14.0$\le$V$\le$21.0), of which $\sim$26,000 (14.0$\le$V$\le$18.0) have $\le$2.5$\%$ photometry suitable for the transit search. All candidates are subjected to a vigorous screening process to identify the very best candidates for further follow-up. The aims of the project are two-fold, firstly to provide a control field for our previous globular cluster transit surveys, both of which produced null results \citep{W2005,W2006}, and secondly to perform a feasibility study for a future wide-field transit survey with the 5.7 deg$^{2}$ ANU SkyMapper telescope \citep{BS2007}.

\begin{figure}[!ht]
\centering
\includegraphics[angle=0,width=7.8cm]{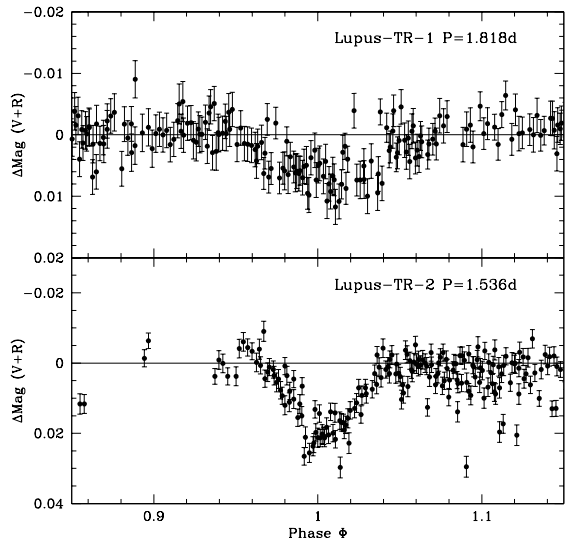}
\caption{Phase-wrapped photometry for Lupus-TR-1 (depth=0.008 mag, duration=3.7hrs) and Lupus-TR-2 (depth=0.02 mag, duration=2.2hrs).\label{tr12}}
\end{figure}

\section{Current Candidates and Follow-up Results}
Analysis of the 2005 dataset identified three transiting systems. The phase-wrapped photometry for the first two can be seen in Fig.\space\ref{tr12}. Both have periods shorter than 2 days, and multiple transits in our data. Low resolution spectra have been taken with the ANU 2.3m telescope and both are consistent with  G dwarf stars. The light curve of Lupus-TR-1 has a 8 mmag transit with a shape typical of a moderately grazing configuration. However, it displays a longer than expected duration and a possible secondary eclipse of $\sim$1mmag depth. 

Lupus-TR-1 was observed in service mode with the 4m AAT and UCLES to derive radial velocities. The results show no discernable velocity variations to our detection limit (230m/s), ruling out a physical binary star. This, coupled with the possible secondary eclipse and unusually long transit duration for a G dwarf star; indicate that this may be a blended system. Lupus-TR-2 displays a short period and moderately V-shaped transits, possibly indicating a grazing configuration. The UCLES radial velocities for this system are currently being analyzed.

\begin{figure}[!ht]
\centering
\includegraphics[angle=0,width=9cm]{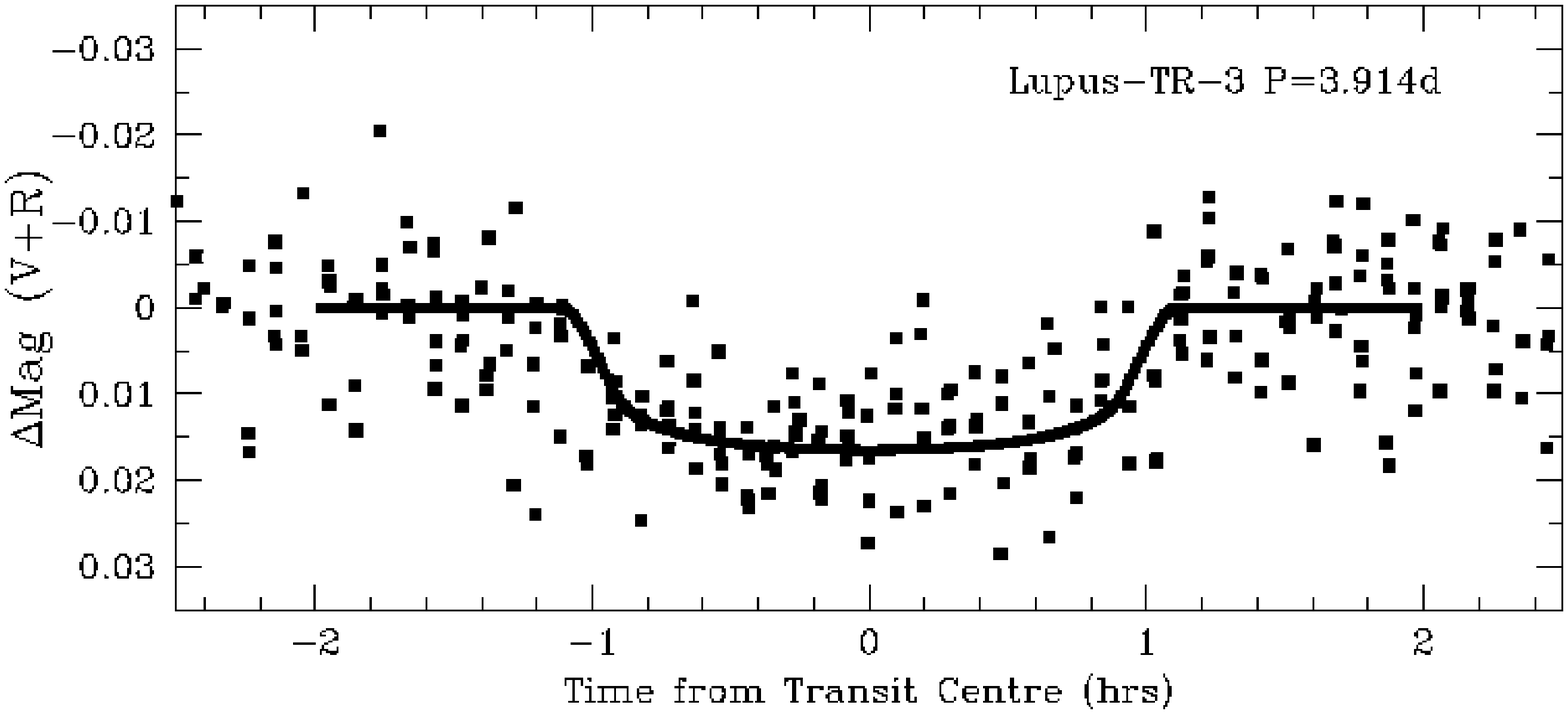}
\caption{The Lupus-TR-3 photometry (P=3.914d), centered on the transit and plotted from the time of transit center in hours. Over plotted is a best-fitting model transit \citep{MA2002} with a resulting depth 16 mmag and total duration of 2.2 hours. We assume HD209458 non-linear limb darkening to describe the ingress and egress phases. The fit to the data is clear.\label{tr3}}
\end{figure}

\section{Lupus-TR-3, a Strong Case for a New Transiting Planet}
The third candidate, Lupus-TR-3, (P=3.914d, V$\sim$16.5) displays a well-sampled flat-bottomed transit, with four fully observed transits and two egresses in our combined 2005 and 2006 data. The phase-wrapped photometry can be seen in Fig.\space\ref{tr3}. The best fitting transit model (over plotted) was produced via the routines of \citet{MA2002} and assumes HD209458 non-linear limb darkening coefficients. This model has a central depth of 16 mmag and a total transit duration of 2.2 hours. From an application of the \citet{TS2005} exoplanet diagnostic (which determines the relative likelihood of transit candidates) Lupus-TR-3 has parameters wholly consistent with a Hot Jupiter planet with $\eta=0.6-0.7$.

From the best-fitting model, the light curve is consistent with a 1.2R$_{\rm{Jup}}$ companion orbiting a 0.98R$_{\odot}$ star, if the companion crosses the average chord length across the stellar disk. If the transit crosses centrally, the solution indicates a 1.0R$_{\rm{Jup}}$ companion to a star of 0.78R$_{\odot}$. We take these values as the upper and lower limits to the companion radius.  A transit of more grazing configuration (hence a larger primary star) would produce a more V-shaped transit. No secondary eclipse or out-of-transit ellipsoidal variations are seen, indicating a companion of negligible luminosity compared to the primary star if the system is not blended. 

The location of the star on the Color Magnitude Diagram of the field as well the stellar proper motion from online USNO and NOMAD1 catalogs are all consistent with a slightly reddened solar-like dwarf star, strengthening the results of the photometric fitting. This system thus appears to harbor a transiting object with radius and period consistent with those of nearby known Hot Jupiters. Multi-band photometry, and low and high resolution spectroscopy are planned to determine conclusively the type of the star and the nature of this promising system.

\acknowledgements
The authors would like to thank Grant Kennedy and Karen Lewis for their assistance with the 2006 observing. 


\end{document}